\documentclass[utf8]{frontiersFPHY} 

\usepackage{url,hyperref,lineno,microtype,subcaption}
\usepackage[onehalfspacing]{setspace}
\usepackage{footnote}

\newcommand{\mee}       {$\langle m_{\beta\beta} \rangle$}

\newcommand{\BBz}       {$0\nu\beta\beta$}

\newcommand{\BBt}       {$2\nu\beta\beta$}
\newcommand{\BB}        {$\beta\beta$}
\newcommand{\Mz}        {$M_{0\nu}$}

\newcommand{\Gz}        {$G_{0\nu}$}

\newcommand{\qval}      {$Q_{\beta\beta}$}                
\newcommand{\Qbb}       {$Q_{\beta\beta}$}

\newcommand{\Tz}        {$T^{0\nu}_{1/2}$}

\newcommand{\Tt}        {$T^{2\nu}_{1/2}$}

\newcommand{\powten}[1] {$10^{#1}$}
\newcommand{\cpowten}[2]{$#1\cdot10^{#2}$}

\newcommand{\nuc}[2]    {$^{#1}$\textrm{#2}} 

\newcommand{\LEG}       {LEGEND}
\newcommand{\Ltwo}      {{\LEG-200}}
\newcommand{\Lthou}     {{\LEG-1000}}

\newcommand{\MJ}        {\textsc{Majorana}}
\newcommand{\DEM}       {\textsc{Demonstrator}}

\newcommand{\be}        {\begin{equation}}
\newcommand{\ee}        {\end{equation}}

\def\keyFont{\fontsize{8}{11}\helveticabold }
\def\firstAuthorLast{Avignone and Elliott} 
\def\Authors{Frank T. Avignone III\,$^{1}$,  and Steven R. Elliott\,$^{2,*}$}
\def\Address{$^{1}$University of South Carolina, Department of Physics,  Columbia , SC, USA \\
$^{2}$Los Alamos National Laboratory, Physics Division, Los Alamos , NM, USA  }
\def\corrAuthor{Corresponding Author}

\def\corrEmail{elliotts@lanl.gov}

\begin{document}
\onecolumn
\firstpage{1}

\title[Ge Detectors and $0\nu\beta\beta$]{The Search for Double Beta Decay with Germanium Detectors: Past, Present and Future} 

\author[\firstAuthorLast ]{\Authors} 
\address{} 
\correspondance{} 

\extraAuth{}

\maketitle

\begin{abstract}
High Purity Germanium Detectors have excellent energy resolution; the best among the technologies used in double beta decay. Since neutrino-less double beta decay hinges on the search for a rare peak upon a background continuum, this strength has enabled the technology to consistently provide leading results. The Ge crystals at the heart of these experiments are very pure; they have no measurable U or Th contamination. The added efforts to reduce the background associated with electronics, cryogenic cooling, and shielding have been very successful, leading to the longevity of productivity. The first experiment published in 1967 by the Milan group of Fiorini, established the benchmark half-life limit $>3\times10^{20}$ yr. This bound was improved with the early work of the USC-PNNL, UCSB and Milan groups yielding limits above $10^{23}$ yr. The Heidelberg-Moscow and USC-PNNL collaborations pioneered the use of enriched Ge for detector fabrication. Both groups also initiated techniques of analyzing pulse waveforms to reject $\gamma$-ray background. These steps extended the limits to just over $10^{25}$ yr. In 2000, a subset of the Heidelberg-Moscow collaboration claimed the observation of double beta decay. More recently, the \MJ\ and GERDA collaborations have developed new detector technologies that optimize the pulse waveform analysis. As a result, the GERDA collaboration refuted the claim of observation with a revolutionary approach to shielding by immersing the detectors directly in radio-pure liquid argon. In 2018, the \MJ\  collaboration, using a classic vacuum cryostat and high-Z shielding, achieved a background level near that of GERDA by developing very pure materials for use nearby the detectors. Together, GERDA and \MJ\  have provided limits approaching $10^{26}$ yr. In this article, we elaborate on the historical use of Ge detectors for double beta decay addressing the strengths and weaknesses. We also summarize the status and future as many \MJ\  and GERDA collaborators have joined with scientists from other efforts to give birth to the LEGEND collaboration. LEGEND will exploit the best features of both experiments to extend the half-life limit beyond $10^{28}$ yr with a ton-scale experiment. 
\tiny
 \keyFont{ \section{Keywords:} double beta decay, neutrino, Ge detectors, Majorana, Dirac}
\end{abstract}

\section{Introduction}
	The very earliest calculation of the rate for two-neutrino double-beta decay (\BBt) is credited to Maria Goeppert-Mayer who predicted the half-life of the decay of \nuc{130}{Te} in 1935~\cite{Goeppert1935}. In 1937, Ettore Majorana built his theory in which neutrinos are their own anti-particles~\cite{Majorana1937}, and in 1939, Wendell Furry proposed testing Majorana's theory by searching for neutrinoless double-beta decay (\BBz)~\cite{Furry1939}. While there were many early efforts to measure double beta decay in the laboratory, the first direct observation of \BBt\ was in \nuc{82}{Se} by Elliott, Hahn and Moe~\cite{Elliott1987} using a Time-Projection-Chamber. The next direct measurements of \BBt\ were made using Ge detectors~\cite{Vasenko1990,Miley1990,Avignone1991}. However, as interesting these experiments were, the most important efforts in building low background Ge detectors were aimed at searching for \BBz\ via the decay, \nuc{76}{Ge} $\rightarrow$ \nuc{76}{Se} + 2e$^-$. In this article we attempt to recall the main highlights of the history of these developments.      

Germanium detectors have many advantages and therefore have provided the most sensitive limits on the \BBz\ half-life (\Tz), and the effective Majorana neutrino mass (\mee), for much of the recent history of neutrino physics. Recently, \nuc{136}{Xe} experiments~\cite{Albert2017,Gando2016} have been more restrictive, although \nuc{76}{Ge} remains highly competitive and the technology is poised to potentially regain its previous supremacy. The search for \BBz\ is fundamentally a search for a rare peak superimposed on a background continuum. Therefore, the excellent energy resolution of Ge detectors, the best of any \BB\ technology, provides highly sensitive discovery potential for the process. This technology also presently has the lowest background when normalized to a resolution-defined region at the \BBz\ Q-value (\qval). The detectors are made from pure metallic Ge resulting in a high atomic density and therefore a relatively large number of atoms per kg of detector. Other benefits include the detectors being mostly insensitive to surface activity and the modest cryogenic requirements of liquid nitrogen temperatures. The technology is well established and has been available as a commercial product for many decades. 

As with any \BB\ technology, all is not ideal with Ge. Relating \Tz\ to \mee\ requires a nuclear matrix element (\Mz), and although \nuc{76}{Ge} benefits from an expectedly high \Mz, \Tz\ also depends on a phase space factor (\Gz) as,

\begin{center}
	(\Tz)$^{-1}$ = \Gz\ \Mz\ \mee$^2$.
\end{center}

\noindent The modest atomic number and \qval\ result in a relatively small \Gz\ compared to other isotopes. It has been calculated by Kotila and Iachello~\cite{Kotila2012} to be \cpowten{2.363}{-15}~/yr and by Stoica and Mirea~\cite{Stoica2013} to be \cpowten{2.34}{-15}~/yr. (For these units of \Gz, \mee\ is taken in units of the electron mass.) Since the \nuc{76}{Ge} \qval\ is low compared to the other most commonly used isotopes and given that \Tz\ scales as $Q_{\beta\beta}^5$, even a small difference can be a significant effect. The enrichment cost has been decreasing but is still a concern. This cost is off-set, however, by the reduced number of detectors that must be fabricated to acquire a given \nuc{76}{Ge} content. It should also be noted that the yearly production of Ge is large compared to the requirements for even a ton-scale experiment, so producing the required isotope will not perturb the economics of the Ge market significantly.  

The long and important history Ge has played in \BB\ has resulted in numerous nuclear physics studies dedicated to the isotope. The calculation of \Mz\ is described elsewhere in this volume and not addressed here. However, one significant example is that of neutron occupancy numbers. These were measured for \nuc{76}{Ge}~\cite{Schiffer2008,Kay09} followed by reconsideration of \Mz\ in light of the additional nuclear structure information. The outcome was that shell model~\cite{Menendez2009a} values increased a bit and the quasi-random phase approximation~\cite{Suhonen2008,Simkovic2009a} results decreased a bit bringing them closer to agreement. Other important nuclear physics measurements include a precise value for \qval\ = 2039.061$\pm$0.007~keV~\cite{Mount2010}, and charge exchange reactions to measure transition strengths~\cite{Thies2012}.

Although we leave a detailed discussion of \Mz\ to others, here we indicate the key references for \nuc{76}{Ge}. The popular nuclear structure models used to calculate \Mz\ are: the interacting boson model (IBM-2)~\cite{Barea2015}, the quasi-particle random phase approximation (QRPA)~\cite{Simkovic2013}, the p-n pairing, QRPA~\cite{Hyvarinen2015}, energy density functional methods (EDF)~\cite{Vaquero2013, Yao2015}, and the interacting shell model (ISM)~\cite{Menendez2009,Horoi2016}. The range of values for \Mz\ for \nuc{76}{Ge} varies from 2.81 to 6.13 for these calculations.

\section{The Germanium Detector and Double-Beta Decay}
Germanium detectors have been the mainstay of nuclear spectroscopy and related fields for more than a half a century. They replaced NaI(Tl) scintillation detectors because the energy resolution is almost 40 times better for $\gamma$ rays with energies near 1 MeV. They consist of single crystals of Ge grown by the Czochralski method~\cite{Kakimoto2013}. Germanium crystals have a diamond structure and Ge has 4 valence electrons. If a Ge crystal has impurities with only 3 valence electrons, then there will be holes throughout the lattice. This is called p-type germanium. In Ge detectors, one or more contact surfaces are heavily doped with lithium to create a surface region of n-type Ge with extra electrons. This configuration constitutes a p-n diode. To operate a Ge detector, a reverse bias voltage is applied which sweeps free holes to the negative contact and conduction-band electrons to the positive contact, essentially clearing the body of the detector of almost all electrical carriers. The crystals are cooled to about 90 K to freeze carriers from thermal excitation to the conduction band. When a $\gamma$ ray, for example, interacts with an electron in the crystal, that electron cascades through the lattice creating electron-hole pairs that migrate towards the opposite sign contacts, creating a displacement current. The carriers reach the electrical contacts, and are detected with a charge-sensitive preamplifier. The number of detected charges is proportional to the energy deposited. 

The early Ge detectors had Li diffused throughout the crystal to create an n-type crystal. These were called GeLi detectors and required cooling at all times to prevent the Li from migrating out of the active volume. GeLi detectors were limited in mass by the ability to drift Li uniformly throughout large crystals. Later, so called {\em intrinsic} or {\em high-purity} Ge detectors were developed in which the natural occurrence of periodic-table, group 3 impurities in the lattice constituted the content of electrical impurities. After zone-refinement and crystallization via the Czochralski technique~\cite{Kakimoto2013}, the electrical impurity level in a typical Ge detector is  \cpowten{(2-3)}{10} electrical impurities per cm$^3$ in the finished detector. This is hyper-pure metal when one considers that there are almost \powten{22} Ge atoms/cm$^3$ in solid germanium.  

The first search for \BB\ using a Ge detector (described below) was by E. Fiorini and his colleagues in the 1960s~\cite{Fiorini1967}. The detector was a 90-gram GeLi detector. Major improvements in technology since then have made searches for \BBz\ far more sensitive. The fabrication of intrinsic Ge detectors, which can have masses of several kilograms, and the use of Ge enriched to 87\% in the candidate parent isotope \nuc{76}{Ge}, from the natural abundance of 7.8\%, led to large sensitivity improvements. Finally, the development of enriched point-contact Ge detectors of about 800 g has revolutionized the ability to discriminate between backgrounds from $\gamma$ rays and \BB\ by pulse shape discrimination. The progress in understanding the origins of background has been substantial. Although the Ge itself is very pure due to the crystal growing process, nearby cables, electronics and shielding may include trace amounts of U/Th. All of these components have seen significant purity improvement.

These developments have resulted in experimental lower bounds on \Tz\ of \nuc{76}{Ge} from the earliest, \cpowten{3}{20} yr to present bounds nearing \powten{26} yr. These recent results have inspired the formation of the \LEG\ Project with the goal of reaching a sensitivity of \Tz\ $\sim$ \powten{28}. This sensitivity would probe the entire inverted neutrino-mass hierarchy for Majorana neutrinos. In this article we discuss the subject from a historical perspective, while also attempting to project into the future.

\section{Early Ge Double-Beta Decay Experiments with Natural Abundance Ge}
In this section we summarize the results of the early experiments culminating in the first use of enriched Ge. We provide the \BBz\ results from detectors fabricated from natural-abundance Ge in Table~\ref{tab:ResultsNatural} and from enriched detectors in Table~\ref{tab:ResultsEnriched}.

\begin{table}[!h]
\caption{Results of \nuc{76}{Ge} \BBz\ experiments using natural abundance Ge. It is interesting to note that in 25 years, the half-life sensitivity of these early experiments increased by more than three orders of magnitude.}
\begin{center}
\begin{tabular}{|l|c|c|c|c|}
\hline\hline
Experiment				&	Year			&		\Tz\ limit (yr)		& Confidence Limit	&	Reference	 			\\
\hline
University of Milan			&	1967			&	\cpowten{3~~}{20}			&	68\%			&	\cite{Fiorini1967}					\\
University of Milan			&	1973			&\cpowten{5~~}{21}			&	68\% 		&	\cite{Fiorini1973}					\\
Battelle-Carolina			&	1983			&\cpowten{1.7}{22}			&	90\% 		&	\cite{Avignone1983}					\\
University of Milan			&	1984			&\cpowten{1.2}{23}			&	68\% 		&	\cite{Bellotti1984}					\\
Guelph, Aptec, Queens		&	1984			&\cpowten{1.5}{22}			&	95\% 		&	\cite{Simpson1984}					\\
Caltech					&	1984			&\cpowten{1.9}{22}			&	68\% 		&	\cite{Forster1984}					\\
Battelle-Carolina			&	1985			&\cpowten{1.4}{23}			&	68\% 		&	\cite{Avignone1985}					\\
Battelle-Carolina			&	1986			&\cpowten{3.0}{23}			&	68\% 		&	\cite{Avignone1986}					\\
UCSB, LBNL				&	1986			&\cpowten{2.5}{23}			&	68\% 		&	\cite{Caldwell1986}					\\
University of Milan			&	1986			&\cpowten{3.3}{23}			&	68\% 		&	\cite{Bellotti1986}					\\
Osaka University			&	1987			&\cpowten{7.3}{22}			&	68\% 		&	\cite{Ejiri1987}					\\
Caltech, Neuch\^{a}tel, PSI	&	1991			&\cpowten{3.4}{23}			&	68\% 		&	\cite{Treichel1991}					\\
UCSB, LBNL				&	1991			&\cpowten{1.2}{24}			&	90\% 		&	\cite{Caldwell1991}					\\
Caltech, Neuch\^{a}tel, PSI	&	1992			&\cpowten{6.0}{23}			&	68\% 		&	\cite{Reusser1992}					\\
\hline\hline
\end{tabular}
\end{center}
\label{tab:ResultsNatural}
\end{table}%

\begin{table}[!h]
\caption{Results of \nuc{76}{Ge} \BBz\ experiments using Ge enriched in \nuc{76}{Ge}. All results are quoted as 90\% CL. In 2001 a subset of the Heidelberg-Moscow collaboration re-analyzed the data claiming evidence for \BBz\ with \Tz\ $=$\cpowten{(0.69-4.18)}{25}~yr (95\% CL)~\cite{Klapdor2004} and has been subsequently refuted by succeeding limits.}
\begin{center}
\begin{tabular}{|l|c|c|c|}
\hline\hline
Experiment				&	Year	&	\Tz\ limit (yr)			&	Reference	 			\\
\hline
ITEP/Yerevan				&	1990	&\cpowten{2.0~}{24}			&	\cite{Vasenko1990}					\\
IGEX-I					&	1994	&\cpowten{1.0~}{24}			&	\cite{Avignone1994b}					\\
IGEX-II					&	1997	&\cpowten{5.7~}{24} 			&	\cite{Aalseth1997}					\\
Heidelberg-Moscow			&	2001	&\cpowten{1.9~}{25}			&	\cite{Klapdor2001}					\\
IGEX-II					&	2002	&\cpowten{1.57}{25}			&	\cite{Aalseth2002}					\\
GERDA-I					&	2013	&\cpowten{2.1~}{25} 			&	\cite{Agostini2013a}\\
\MJ\						&	2018	&\cpowten{2.7~}{25} 			&	\cite{Guiseppe2018}\\
GERDA-II					&	2018	&\cpowten{8.0~}{25} 			&	\cite{Agostini2018}\\
\hline\hline
\end{tabular}
\end{center}
\label{tab:ResultsEnriched}
\end{table}%

\subsection{The First University of Milan Experiments}
\begin{figure}[h!]
\begin{center}
\includegraphics[width=12cm]{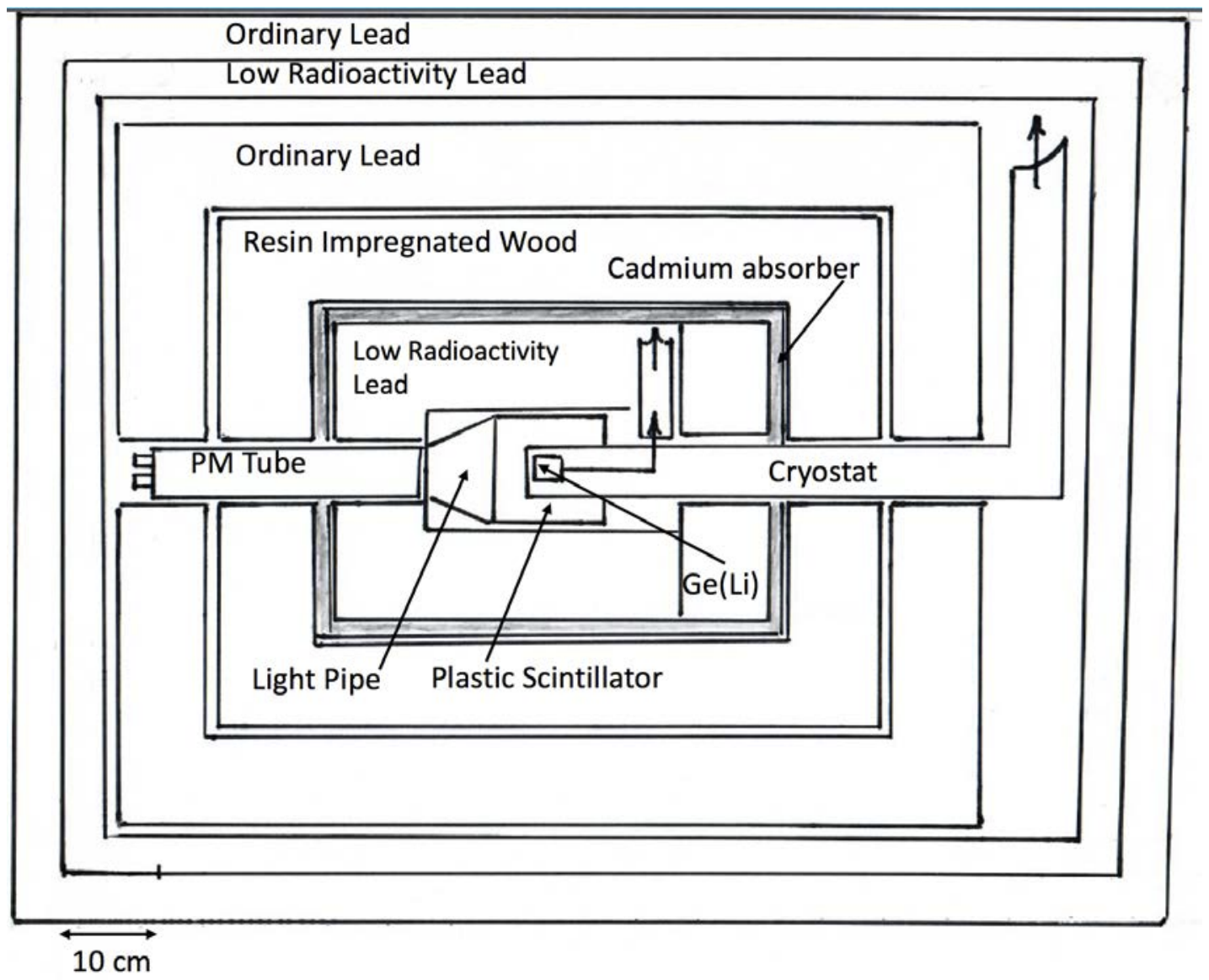}
\end{center}
\caption{A diagram of the Milan Mont-Blanc Tunnel Double-Beta Decay Experiment.}
\label{fig:MilanLayout}
\end{figure}

The first search for \BBz\ of \nuc{76}{Ge} was performed by Fiorini and his University of Milan colleagues~\cite{Fiorini1967}. While many of the later results were from the analyses of data from low-background Ge counting facilities, the Milan experiment was built for the express purpose of testing lepton conservation. The heart of apparatus consisted of a 17 cm$^3$ ($\sim$90 g) GeLi detector, with an energy resolution of $\sim$4.7~keV at 1.32~MeV. (See Fig.~\ref{fig:MilanLayout}.) The detector was surrounded on all sides, except for the end, by a plastic scintillator veto. The entire apparatus was surrounded by a shield of 10~cm of low background lead, surrounded by a thin cadmium neutron absorbing shield, encased in a 10-cm thick box of resin impregnated wood as a neutron moderator. The outer shield was 10~cm of ordinary lead. The experiment was located in the Mount Blanc Tunnel at a location with 4200 meters of water equivalent (mwe) overburden. The background data were taken for 712~h of live time. The background rate at \qval\ was \cpowten{1.1}{-2} counts/(keV h), which in today's terminology is \cpowten{1.06}{3} counts/(keV kg yr). The data implied a bound of \Tz\ $\geq$ \cpowten{3.1}{20} yr (68\% CL). In addition to being the first high-resolution search for \BBz, this was the first experiment in which the source and detector were one and the same, yielding an excellent detection efficiency.

In 1973, the Milan group published their results from a greatly upgraded experiment located in the same location in the Mount Blanc Tunnel~\cite{Fiorini1973}. The detector in this case was a GeLi detector with an active volume of 68.5~cm$^3$ ($\sim$365 g). In this shield, the plastic scintillator was eliminated because it brought background. Immediately surrounding the detector cap was a Nylon-Marinelli beaker filled with doubly-distilled Hg. This was surrounded with 4 cm of electrolytic copper, encased in 10~cm of low background lead, followed by 10~cm of ordinary lead. The outer lead shield was  surrounded by a 2-mm thick cadmium sheet and the entire shield was then enclosed by 20~cm of paraffin to moderate background neutrons. 

There were two data collection periods totaling 4400 h of live time. The background was \cpowten{4.3}{-3} counts/(keV h). This is equivalent to $\sim$\cpowten{1.02}{2} counts/(keV kg yr), a factor of ten improvement in background over the 1967 result. The final result was \Tz\ $\geq$\cpowten{5.0}{21} yr (68\% CL)~\cite{Fiorini1973}. The Milan Group built a new experiment with two intrinsic Ge detectors of fiducial volumes of 117 cm$^3$ and 138 cm$^3$, in a common shield~\cite{Bellotti1986}. There were several improvements in low background construction materials in the cryostat and shielding. There were two counting periods in which the shielding configuration underwent minor changes. The total counting time was 1.76~yr, and the resulting limit was \Tz\ $\geq$\cpowten{3.3}{23}~yr (68\% CL).

\subsection{The Early Battelle-Carolina Experiments}
	The field of experimental \BBz\ was dormant for a while after the 1973 Milan result. Renewed interest in \BBz\ was driven by several events. First, Lubimov claimed that the electron neutrino had a mass of above 14 eV, from the data of the ITEP tritium-end-point experiment~\cite{Lubimov1980}. Second,  interest in the theory of Grand Unification was intense and third, the shell-model calculations of the nuclear matrix elements by Haxton, Stevenson and Strottman~\cite{Haxton1981}, indicated significant strength for the  decay of \nuc{76}{Ge}. With these new motivations, Avignone and Greenwood proposed in 1979 an experiment, based on a Monte Carlo study with a high-purity Ge detector enclosed in a NaI(Tl) Compton suppression shield~\cite{Avignone1979}. The assumed backgrounds in this proposal were taken to be similar to the Milan experiments discussed above. A trial of the experiment suggested in Ref.~\cite{Avignone1979} was proposed to the team of Brodzinsky and Wogman at the Battelle Pacific Northwest Laboratory (now PNNL). For several years the Battelle-Carolina Collaboration worked on improving the backgrounds due to the construction materials in  copper cryostats. The intrinsic Ge detector had an active volume of 125~cm$^3$. It was operated inside a two-inch thick NaI(Tl) Compton suppression shield, inside a lead shield, covered by a boron-loaded polyethylene neutron shield, with a plastic cosmic-ray shield above the entire apparatus. The experiment was operated above ground for a live time of 4054 h, resulting in the bound \Tz\ $\geq$\cpowten{1.7}{22}~yr~\cite{Avignone1983}. The background rate was similar to previous experiments at \cpowten{1.04}{2} counts/(keV kg yr). The detector was then moved to a location at 4850~ft below the surface in the Homestake Gold Mine in Lead, South Dakota, in part of the Solar Neutrino Laboratory of Raymond Davis. That location has an overburden of $\sim$4300 mwe~\cite{Heise2015}. The detector was housed in a 40-cm thick, ordinary lead shield. The energy resolution was 3.7~keV at  \qval\ and the background rate was 47~counts/(keV kg yr).  The detector was operated for 8089 h at the same site in the Homestake mine with the result, \Tz\ $\geq$\cpowten{1.4}{23}~yr (90\% CL)~\cite{Avignone1986}. The construction details are given in Ref.~\cite{Brodzinski1990}. The lesson learned was that much of the background, although reduced, was coming from the cryostat itself. This fact led to a significant R\&D effort by the Batelle-Carolina Collaboration to create ultra-low background copper by electroforming from CuSO$_4$ solutions onto stainless steel mandrels. The results were the production of all six of the cryostats for the International Germanium Experiment, IGEX, with electroformed copper.  The IGEX experiments are discussed below.

\subsection{The Guelph, Aptec, Queens Experiment}
	At a time shortly after the 1983 Battelle-Carolina experiment, the team of J.J. Simpson was operating a commercially built, low background Ge detector underground. The intrinsic 208~cm$^3$ ($\sim$1.1~kg) Ge detector was operated in a salt mine near Windsor, Ontario, at a depth of about 330~m~\cite{Simpson1984}. The detector was shielded with 20~cm of lead. In the final run, a 6-mm thick mercury shield was placed inside the lead castle, which absorbed the low energy bremsstrahlung from the decay of \nuc{210}{Bi}, a daughter of the 22-yr \nuc{210}{Pb} in the shield. Although the lead was between 150 and 200 yr old, this radiation still remained, demonstrating that the lead of the shield contained a high level of \nuc{238}{U}. The detector operated for 2363 h. The result was a bound of \Tz\ $\geq$\cpowten{3.2}{22}~yr (68\% CL), or \Tz\ $\geq$\cpowten{1.5}{22}~yr at (95\% CL).
	
\subsection{The Caltech and the Neuch\^{a}tel-Caltech-PSI Experiments}
\begin{figure}[h!]
\begin{center}
\includegraphics[width=12cm]{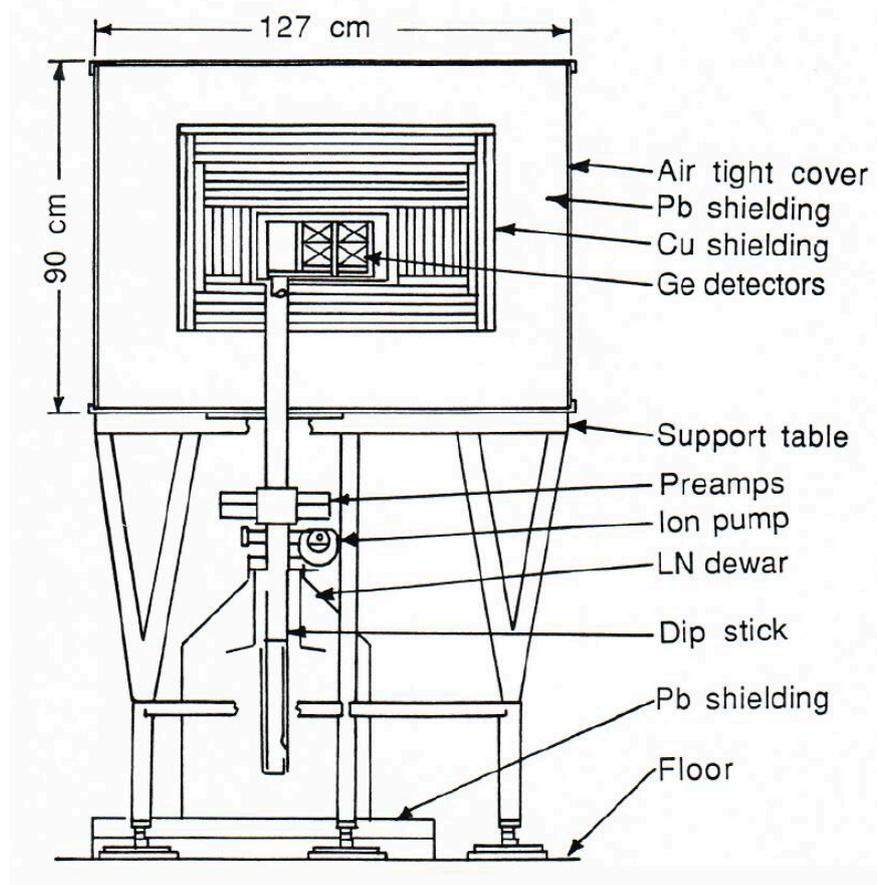}
\end{center}
\caption{A diagram of the Caltech, Nuch\^{a}tel, PSI experiment with four natural abundance, high-purity Ge detectors.}
\label{fig:PSILayout}
\end{figure}

	The Caltech Group began their experimental series by setting up a shielded detector above ground in a sub-basement at Caltech. The overburden was only 3~mwe. The Princeton Gamma-Tech, high-purity coaxial Ge detector had a $\sim$90~cm$^3$ fiducial volume. The detector was surrounded with 15~cm of electrolytic copper, followed by 15~cm of lead. The shield was enclosed in an airtight box to protect the detector from airborne radon. The final result from 3820~h of this essentially above-ground experiment was \Tz\ $\geq$\cpowten{1.9}{22}~yr (68\% CL)~\cite{Forster1984}. The next experiment involving the Caltech group was in collaboration with the University of Neuch\^{a}tel and Paul Scherrer Institute~\cite{Treichel1991}. It involved 8 high-purity 140-cm$^3$ Ge detectors, with a combined volume of 1095~cm$^3$ or 5.83~kg. The array was operated in the Gotthard Tunnel with an overburden of 3000~mwe. The array of detectors was surrounded with 15~cm of oxygen-free, high-conductivity (OFHC) copper, followed by 18~cm of lead, all contained in an aluminum radon shield. (See Fig.~\ref{fig:PSILayout}.) The array was operated for 6.2~kg~yr of live time with resulting limits of \Tz\ $\geq$\cpowten{2.0(3.4)}{23}~yr 90\%(68\%) CL.  The final report of this collaboration~\cite{Reusser1992} reported  \Tz\ $\geq$\cpowten{6.0}{23}~yr (68\% CL) from 10.0~kg~yr. This was the strongest bound from the natural-abundance Ge detectors.
	
\subsection{The UCSB, LBNL Experiments}
	The UC Santa Barbara, Lawrence Berkeley National Laboratory experiment began with two intrinsic Ge detectors of 178~cm$^3$ and 158~cm$^3$ operating above ground in a NaI(Tl) Compton-suppression shield for 1618~h. A later version was a configuration of four intrinsic detectors with a total fiducial volume of 658~cm$^3$. The array had new interesting construction features, for example Si cold fingers to avoid the background due to the copper commonly used~\cite{Caldwell1986}. The array was operated for 3550~h, 200~m below ground in the Power Station in the Oroville Dam in Northern California. The final result was \Tz\ $\geq$\cpowten{2.5}{23}~yr (68\% CL). The array was later used to produce very interesting data in the search for Cold Dark Matter~\cite{Caldwell1988}.
	
\subsection{The Osaka University Experiment}
	The first phase of the Osaka experiment began above ground with a 171~cm$^3$ intrinsic Ge detector in a $4\pi$ NaI(Tl) Compton suppression shield, surrounded by a mercury shield~\cite{Ejiri1987}. The detector was operated for 1600~h in the Kamioka Underground Laboratory, with an overburden of 2700 mwe. In the second phase, the detector was operated for 7021~h, but without the mercury shield. The final result was \Tz\ $\geq$\cpowten{7.3}{22}~yr (68\% CL).
	
\subsection{The ITEP-Yerevan Experiment, and the Early Measurements of \BBt\ of \nuc{76}{Ge}}
\begin{figure}[h!]
\begin{center}
\includegraphics[width=12cm]{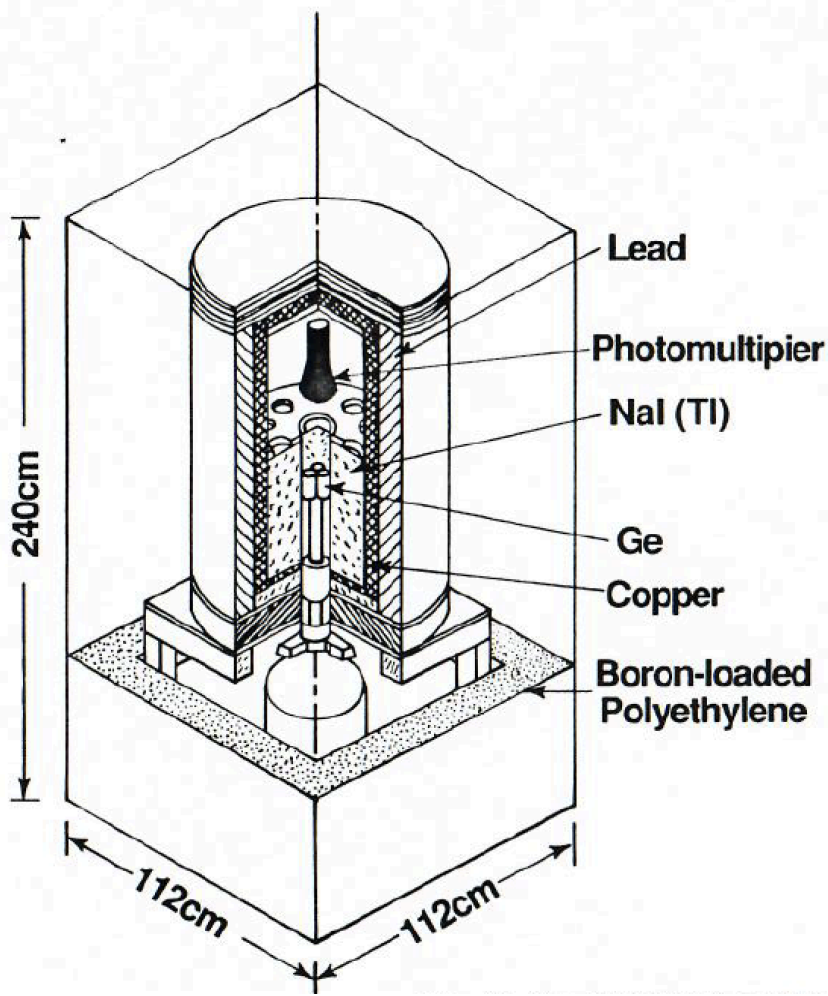}
\end{center}
\caption{A diagram of the ITEP/Yerevan experiment showing the three crystals, two enriched, and one of natural abundance.}
\label{fig:ITEPLayout}
\end{figure}

	This experiment was the first search for \BBz\ of \nuc{76}{Ge} with detectors fabricated with Ge enriched in \nuc{76}{Ge}~\cite{Vasenko1990}. This experiment consisted of three GeLi detectors, two of which were fabricated with Ge enriched to 85\% in \nuc{76}{Ge}. (See Fig.~\ref{fig:ITEPLayout}.) The total mass of \nuc{76}{Ge} was 1008~g. The three crystals were on the end of a vertical cold finger inside of a NaI(Tl) Compton shield surrounded by several cm of copper followed by lead. The entire apparatus was inside a boron-loaded Polyethylene box 112~cm$\times$112~cm$\times$240~cm high. The experiment was operated 245~m underground, in the Avansk Mine in Yerevan, Armenia. The background at \qval\ was 2.5~counts/(keV kg yr) for the two enriched crystals and 2.1~counts/(keV kg yr) for the natural crystal. A final analysis of the results yielded a limit of \Tz\ $\geq$\cpowten{1.0}{24}~yr (90\%). 

	In addition, this experiment was the first direct observation of the \BBt\ decay of \nuc{76}{Ge}, and only the second such laboratory measurement following that in \nuc{82}{Se}~\cite{Elliott1987}. The result of the ITEP-Yerevan experiment was \Tt\ $=$\cpowten{(9\pm1)}{20}~yr. This result was submitted to Modern Physics Letters on 23 April 1990~\cite{Vasenko1990}. Later that year, the Battelle-Carolina group submitted a similar result: \Tt\ $=$\cpowten{1.1^{+0.6}_{-0.3}}{21}~yr (95\% CL), to Physical Review Letters from data taken with two 1.05-kg, ultra-low background, natural abundance, intrinsic Ge detectors~\cite{Miley1990}. The two groups then merged and placed one of the ITEP-Yerevan enriched GeLi detectors in the Battelle-Carolina Cryostat to re-measure the half life. The result was \Tt\ $=$\cpowten{(9.2^{+0.7}_{-0.4})}{20}~yr (2$\sigma$)~\cite{Avignone1991}. It was later demonstrated that all three of these results were contaminated with internal radioactivity generated by spallation reactions of hard cosmic-ray neutrons (e.g. \nuc{60}{Co}, \nuc{65}{Zn},  \nuc{68}{Ge}). These backgrounds produced events, which were partially attributed to \BBt, resulting in deducing shorter half-lives. Results by IGEX presented at ERICE in 1993~\cite{Avignone1994}, corrected for these backgrounds and found \Tt\ $=$\cpowten{(1.27^{+0.21}_{-0.16})}{21}~yr (1$\sigma$). Later experiments demonstrated that these corrections for internal background, while in the correct direction, were still inadequate. Historically, the value for \Tt\ has increased for \nuc{76}{Ge} indicating that the background subtraction is very difficult. Table~\ref{tab:ResultsTwoNu} summarizes the measurements of \Tt.

	\begin{table}[!h]
\caption{Results of \nuc{76}{Ge} \BBt\ experiments. The half life continues to creep up due to the complexity of subtracting the background. As experiments improve and the background is either reduced or better fit, \Tt\ increases.}
\begin{center}
\begin{tabular}{|l|c|l|c|c|}
\hline\hline
Experiment				&	Year	&	\Tt\ measurement (\powten{21} yr)					&Confidence 		&	Reference	 			\\
						&		&												& Level			&		 			\\
\hline
ITEP/Yerevan				&	1990	&	$0.9\pm0.1$ 									&	68\% 		&	\cite{Vasenko1990}					\\
Batelle-Carolina			&	1990	&	$1.1^{+0.6}_{-0.3}$								&95\% 			&	\cite{Miley1990}					\\
ITEP-Yerevan/Batelle-Carolina	&	1991	&	$0.92^{+0.07}_{-0.3}$							 &	90\% 		&	\cite{Avignone1991}					\\
IGEX					&	1994	&	$1.27^{+0.21}_{-0.16}$		 					&	68\% 		&	\cite{Avignone1994}					\\
Heidelberg-Moscow			&	1997	&	$1.77\pm0.01(stat)^{+0.13}_{-0.11}(syst)$				&68\% 			&	\cite{Gunther1997}		\\
IGEX					&	1999	&	$1.45\pm0.20$									&68\%			&	\cite{Morales1999}					\\
Heidelberg-Moscow			&	 2001& 	$1.55\pm0.01(stat)^{+0.19}_{-0.15}(syst)$				&68\% 			&	\cite{Klapdor2001}					\\
Heidelberg-Moscow			&	2003	&	$1.74^{+0.18}_{-0.16}$							&68\% 			&	\cite{Dorr2003}					\\
Heidelberg-Moscow			&	2005	&	$1.78\pm0.01(stat)^{+0.07}_{-0.09}(syst)$				&68\% 			&	\cite{Bakalyarov2005}					\\
GERDA					&	2013	&	$1.84^{+0.14}_{-0.10}$							&68\% 			&	\cite{Agostini2013} \\
GERDA					&	2015	& 	$1.926\pm0.094$								&68\% 			&	\cite{Agostini2015} \\
\hline\hline
\end{tabular}
\end{center}
\label{tab:ResultsTwoNu}
\end{table}%

\subsection{The International Germanium Experiments (IGEX): The First Enriched-High-Purity Ge Detectors}
\begin{figure}[h!]
\begin{center}
\includegraphics[width=12cm]{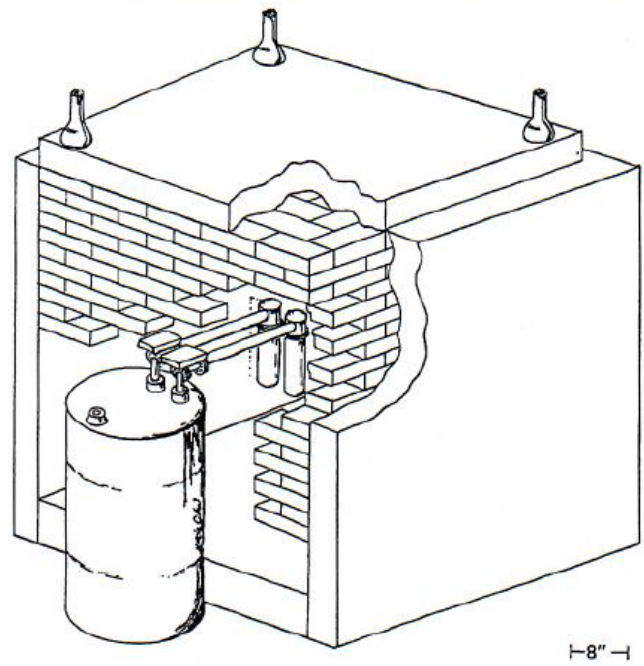}
\end{center}
\caption{A diagram of the first IGEX-II detectors in the Homestake gold mine showing the detectors hanging vertically.}
\label{fig:IGEXLayout}
\end{figure}

	In 1988, the Battelle-Carolina Collaboration concentrated on lowering background by electroforming the cryostat parts from CuSO$_4$ solution, and acquiring Ge enriched to 86\% in \nuc{76}{Ge}. A collaboration was formed between Battelle Northwest, the University of South Carolina, the Institute of Theoretical and Experimental Physics (ITEP) Moscow, the Institute of Nuclear Research (INR) Moscow, and the University of Zaragoza. Over several installments, a total of 18~kg of  Ge, enriched to 86\% in \nuc{76}{Ge} were imported to the U.S. from the two Russian institutes in oxide form. The first 5~kg from INR was reduced and zone refined by Mr. James Meyer at Eagle Picher Inc. Three 190~cm$^3$ high-purity Ge detectors were fabricated and tested in the Homestake gold mine, at the 4850-ft level. (See Fig.~\ref{fig:IGEXLayout}.) While the energy resolution and general operation of the detectors was excellent, measurements determined that the fiducial volumes were only about 135~cm$^3$. Difficulties in crystal growth required the Li deposition on the outer surfaces to be thicker than normal.  These three detectors constituted IGEX-I. One was operated in the Homestake gold mine, one in the Canfranc Underground Laboratory, in Canfranc Spain (1380~mwe), and one in the Baksan Neutrino Observatory, in Russia (660~mwe). The first results from IGEX-I were presented at the International Conference on Topics in Astroparticle and Underground Physics (TAUP-93), at Laboratori Nazionali del Gran Sasso (LNGS), in Assergi, Italy~\cite{Avignone1994b}. The data from the three detectors were combined with the result \Tz\ $\geq$\cpowten{1.0}{24}~yr (90\% CL). The average background was 0.3~counts/(keV kg yr). It was also announced at that meeting that the first of three IGEX-II detectors had been fabricated and tested. It had a fiducial volume of $\sim$400~cm$^3$, and an energy resolution of 2.16~keV FWHM at 1332~keV. The first IGEX results using pulse-shape discrimination to identify background events from $\gamma$ rays, was presented at Neutrino-96 at Helsinki. The result from 34.4~mole~yr of data was \Tz\ $\geq$\cpowten{5.7}{24}~yr (90\% CL)~\cite{Aalseth1997}. While IGEX was first to build and operate high-purity Ge detectors enriched in \nuc{76}{Ge}, by the time of this meeting, the Heidelberg Moscow Collaboration was already operating $\sim$400~cm$^3$ high-purity detectors and had excellent results (discussed below). The IGEX technique for pulse-shape discrimination was described in detail with IGEX-I detectors in Ref.~\cite{Aalseth1998}, and later using the larger IGEX-II detectors in Ref.~\cite{Gonzales2003}.
	
	During the period 1996 and 1997, the IGEX collaboration had three high-purity enriched coaxial detectors produced with active volumes of $\sim$400~cm$^3$. The IGEX detectors had a unique configuration hanging from the end of the cold fingers. The cold finger rose from the liquid nitrogen bottle, made a 90$^{\circ}$ turn to horizontal, extended through the shield to the cold plate from which the detector cryostats were hung vertically down. This configuration prevented the radioactive contamination of Xeolite cryopump material from having a direct line of sight to the detector. The three IGEX-II detectors were tested at Homestake, then carried by ship to Barcelona Spain, and installed in the Canfranc Underground Laboratory of the University of Zaragoza. It is important to point out that by this time, the experiment of the Heidelberg-Moscow group was operating four large enriched detectors in the LNGS, and exceeded IGEX in exposure.  
	
	While there were a number of IGEX updates published in conference proceedings, the first publication of results, including the data taken with the IGEX-II detectors was in 1999, based on 78.84~mole~yr of exposure. The total mass of detectors was 8.1~kg. The resulting bound was \Tz\ $\geq$\cpowten{0.8}{25}~yr (90\% CL). The data were subjected to the pulse-shape discrimination techniques described in Refs.~\cite{Aalseth1998} and \cite{Gonzales2003}. The final IGEX result was published after a total of 117~mole~yr of exposure: \Tz\ $\geq$\cpowten{1.57}{25}~yr (90\% CL)~\cite{Aalseth2002}. The publication of this final IGEX result set a controversy in motion. A subset of the Heidelberg-Moscow Collaboration claimed that serious errors were made in the analysis of the final IGEX results~\cite{Klapdor2004}. The response by the IGEX collaboration~\cite{Aalseth2004} clearly justified the IGEX analysis and the final result given in Ref.~\cite{Aalseth2002}.

\subsection{The Heidelberg-Moscow Experiment}
	The Heidelberg-Moscow Collaboration launched a very impressive experiment with five coaxial-high-purity Ge detectors enriched to 88\% in \nuc{76}{Ge}, with a total mass of 11.5~kg, and an active volume with 10.96 kg, operating in LNGS. The laboratory has an overburden of about 3500~mwe. The detectors were enclosed in a shield with a 10-cm inner layer of ultrapure lead, surrounded by 20~cm of pure Boliden lead, enclosed in a metal box flushed with high-purity nitrogen. The shield was surrounded by 10~cm of boron-loaded polyethylene~\cite{Balysh1995,Baudis1997,Klapdor1997}. The experiment had an effective pulse shape analysis technique for identifying and removing background events~\cite{Helmig2000}. It operated from 1990-2003 with a total exposure of 71.71 kg y. It was the most sensitive \nuc{76}{Ge} experiment until the GERDA experiment commenced. There were many publications presenting the results over the years. In 2001 the collaboration published the best bound on  decay:  \Tz\ $\geq$\cpowten{1.9}{25}~yr (90\% CL)~\cite{Klapdor2001b}. Later that year, a subset of the collaboration published a claim of direct observation of \BBz\ of \nuc{76}{Ge}, with a half-life of \Tz\ $=$\cpowten{(0.8-18.3)}{25}~yr (95\% CL), based on 46.5~kg~yr of exposure~\cite{Klapdor2001b,Klapdor2002}. The final range of claimed values for the discovery, \Tz\ $=$\cpowten{(0.69-4.18)}{25}~yr (95\% CL), and the entire history of these experiments from 1990 to 2003, is given in Ref.~\cite{Klapdor2004}. 

	The claim of discovery was critiqued in an article coauthored by a broad list of authors~\cite{Aalseth2002b}, and later excluded by results from the GERDA Experiment (discussed below). This claim has also been excluded by the Xe experiments (KamLAND-Zen~\cite{Gando2016} and EXO~\cite{Albert2017}), but the direct comparison between Ge experiments removes any caveats regarding the relative matrix element values. In addition to the search for \BBz, the collaboration measured \Tt\ = \cpowten{(1.55\pm0.01(stat)^{+0.19}_{-0.15}(syst))}{21} yr ~\cite{Klapdor2001} followed later by \cpowten{(1.74^{+0.18}_{-0.16})}{21}~yr~\cite{Klapdor2004c}.

\section{Modern Day Double-Beta Decay Experiments}
The Heidelberg-Moscow experiment was based on 13 yr of very low background operation. Hence it would be a very difficult experiment to repeat. Furthermore during the 1990's and into the 2000's, \BB\ experiments took a back seat to the interest in solar neutrino experiments due to their role in $\nu$ oscillations. As a result there was another hiatus in \BBz\ results. Interest built for \BB\ again in the late 2000's due to the claim of an observation and the confirmation that $\nu$ oscillations exist and, by inference, massive light neutrinos exist. In addition, the $\nu$ physics parameters indicated by the oscillation results meant that new \BBz\ experiments would have discovery potential for a significant range of possible \mee\ values. The field of \BB\ saw a large number of new proposals advance by about 2010, including those based on \nuc{76}{Ge}.

One key development in Ge detector technology has greatly improved their pulse shape analysis capability. That is the use of a point-contact. Originally developed for their low capacitance~\cite{Luke1989}, it was after the development of modern-day transistors that the full power of this detector design began to be exploited, in particular for dark matter experiments~\cite{Barbeau2007}. The advantage for \BB\ arises because the weighting potential is strongly peaked at the contact for this geometry. This results in an electronic signal that predominately forms only when drifting charge nears the contact. Therefore, an event with multiple energy deposits within a detector will have pulse shape distinct from that of a single-site energy deposit.  As \BB\ is a single-site energy deposit and many backgrounds are multiple site events, this is a powerful rejection capability and point-contact detectors substantially surpass the performance of the semi-coax Ge detector design that had been the field's workhorse. \MJ\ and GERDA~\cite{Budjas2009} further developed and used this technology to great success.

During research and development for the \MJ\ and GERDA programs, the use of segmented detectors was considered.  Segmented detectors provide enhanced waveform analysis and hence improved background rejection. A number of studies~\cite{Elliott2006,Abt2007,Abt2007a,Abt2007b} were done considering the added advantages of segmentation on the reduction of background versus the disadvantages of the extra complexity and background due to the additional electronic channels and cables. The \MJ\ collaboration successfully developed a segmented enriched detector~\cite{Leviner2014} that showed some promise. After the development of point-contact detectors, however, it became clear that the advantages of segmentation were outweighed by the disadvantages. Segmented detectors for \BB\ were not further pursued.

\subsection{The \MJ\ Experiment}
The \MJ\ \DEM~\cite{Abgrall2014,Aalseth2018} experiment was established to demonstrate that backgrounds can be controlled to a level that would justify a large (ton scale) \nuc{76}{Ge} effort. Previous \nuc{76}{Ge} experiments with compact, high atomic-number shielding indicated that the classic design of a vacuum cryogenic-cryostat filled with Ge detectors surrounded by Pb could extend the reach of \BB\ physics. The \MJ\ project, named in honor of Ettore Majorana and based on this concept, began construction in 2010 with initial commissioning data collected in 2015.

\begin{figure}[h!]
\begin{center}
\includegraphics[width=12cm]{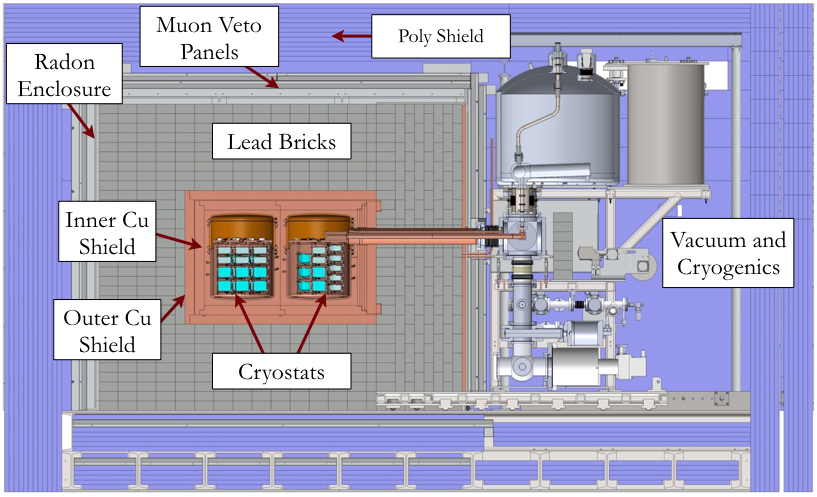}
\includegraphics[width=6cm]{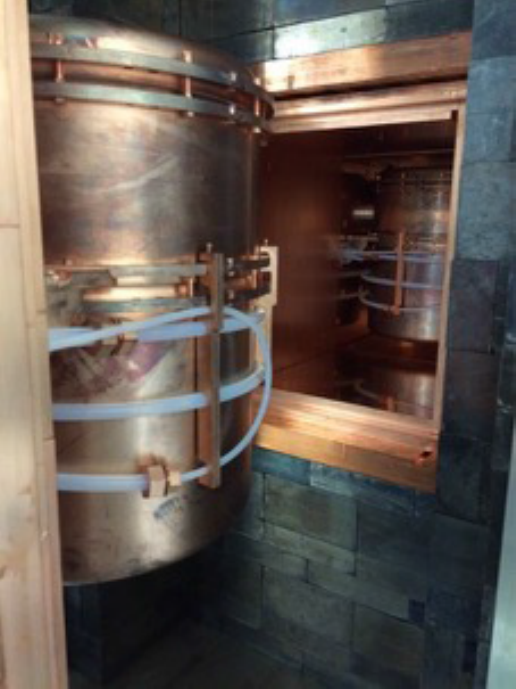}
\end{center}
\caption{ Left: The shield concept for the \MJ\ \DEM. Right: A photograph of one cryostat ready for insertion into the shield with the other, already installed, visible in the background. Figure and photo courtesy of the \MJ\ Collaboration.}\label{fig:MJDLayout}
\end{figure}

The ongoing experiment is sited 4300 mwe underground at the 4800-ft level of the Sanford Underground Research Facility (SURF)~\cite{Heise2015}. The Ge detectors, 44.1 kg total with 29.7 kg enriched to 88\% in \nuc{76}{Ge}, are enclosed within two electroformed-Cu~\cite{Hoppe2014} cryostats. The detectors are mounted in groups of 3 to 5 and hung  as strings from a cold plate cooled by a thermosyphon~\cite{Aguayo2013a}. Very low radioactivity, front-end electronic boards~\cite{Barton2011}, placed very close to the detectors, maintain signal fidelity while providing the initial amplification stage. The cryostat is contained within a 5-cm thick electrofromed Cu layer, a 5-cm thick commercial C10100 copper layer, a 45-cm thick Pb shield, two layers of plastic-scintillator cosmic-ray veto panels, 5 cm of borated poly and finally 25 cm of high density polyethylene. The material inside the veto layer is contained in an Al box that is purged with boil-off N$_2$ to displace Rn-laden room air. (See Fig.~\ref{fig:MJDLayout}.) All materials comprising the experiment were analyzed for their radiopurity~\cite{Abgrall2016}. The processing of Ge for \MJ\ developed recycling techniques~\cite{Abgrall2018} that are critical to reduce the amount of required raw material to fabricate a given mass of detectors.

\begin{figure}[h!]
\begin{center}
\includegraphics[width=8cm]{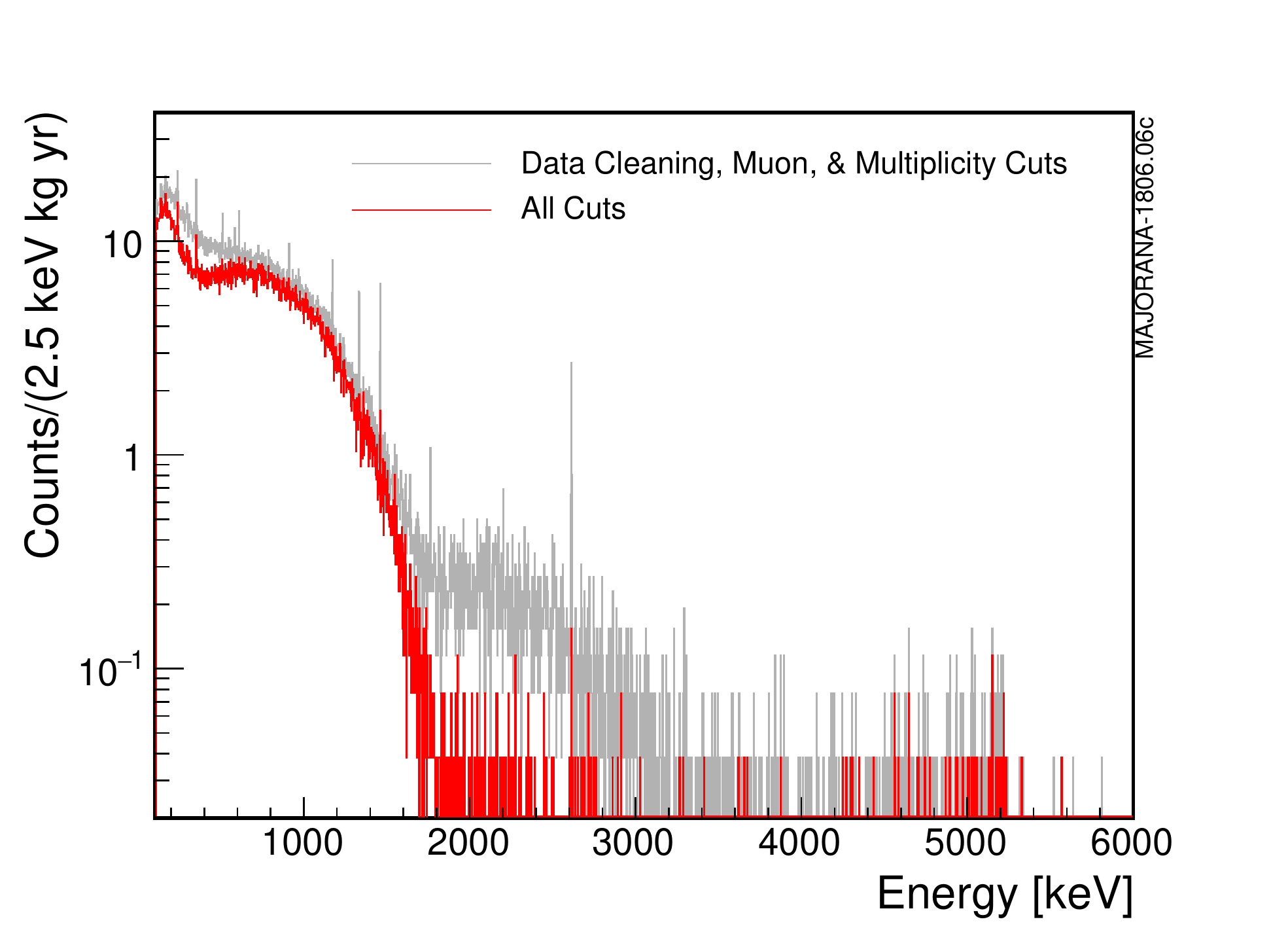}
\includegraphics[width=8cm]{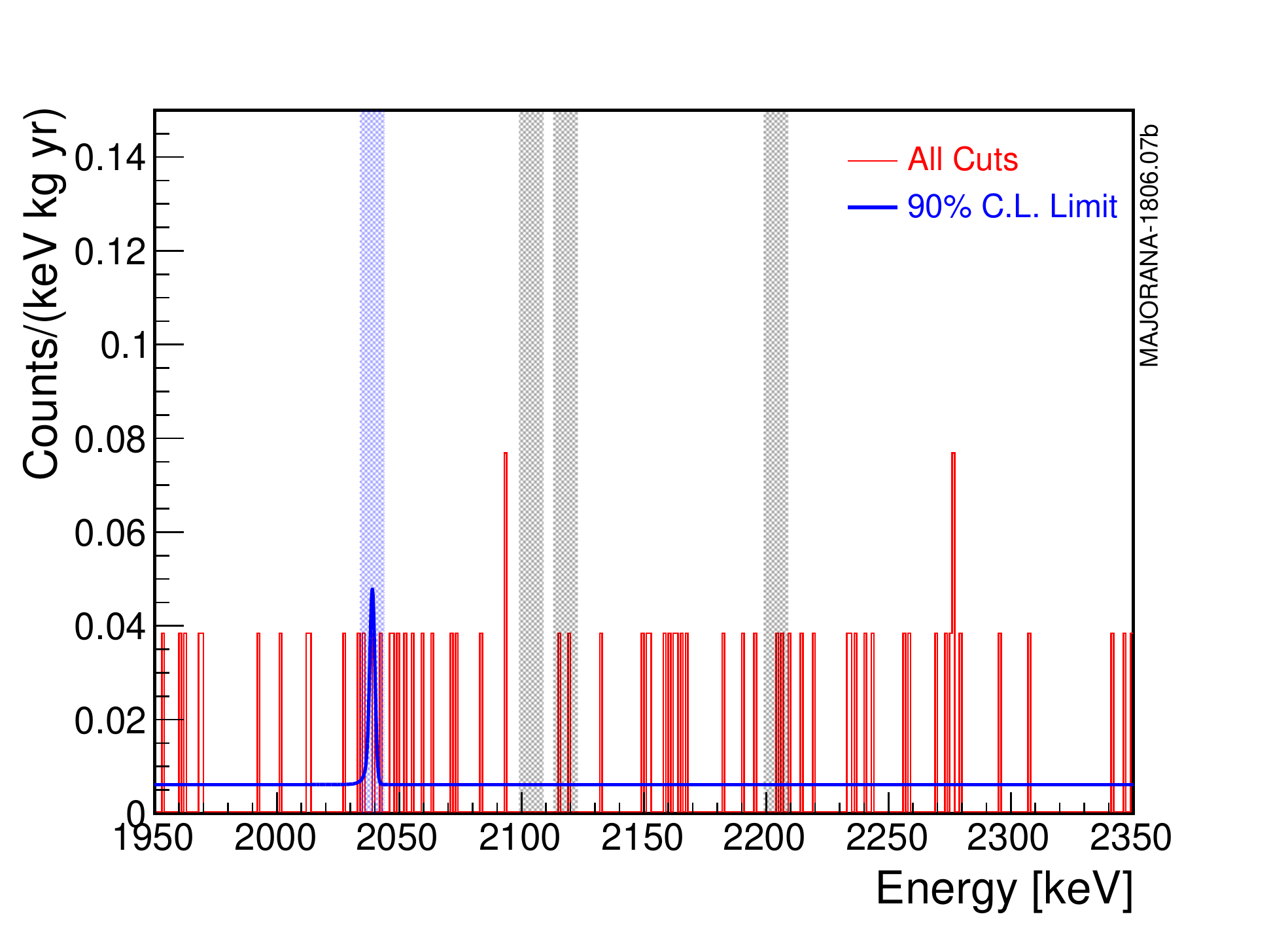}
\end{center}
\caption{Left: the whole spectrum from the \MJ\ \DEM\ from 26 kg-yr of exposure. Right: The specrum near the \Qbb\ for Ge at 2038 keV~\cite{Guiseppe2018}. Figures courtesy of the \MJ\ Collaboration.}\label{fig:MJDspectra}
\end{figure}

Initial results from the \DEM\ were based on an exposure of 10 kg yr~\cite{Aalseth2018}. A second data release~\cite{Guiseppe2018} based on 26 kg yr of exposure yielded a half-life limit of \cpowten{>2.7}{25} yr (90\% CL). After removal of non-physical events, events in coincidence with the muon veto, events with multiple detectors in coincidence, and pulse shape analysis to remove single-crystal events with multiple energy deposits and surface $\alpha$ interactions, the final background is $11.9\pm2.0$ counts/(FWHM t yr) or \cpowten{(4.7\pm0.8)}{-3} counts/(keV kg yr) from the 21.3 kg yr lowest background configuration. The spectra from the full 26 kg yr exposure are shown in Fig.~\ref{fig:MJDspectra}. The energy resolution, 2.5 keV FWHM at \Qbb, is the best achieved of any \BB\ experiment. Although the analysis is not yet complete, early studies indicate the dominate source of background in the \DEM\ is not from nearby components within the detector arrays~\cite{Caldwell2018}.

The low background, excellent energy resolution and low energy threshold permit a variety of other physics measurements with \MJ, including tests of the Pauli Exclusion Principle, electron decay, bosonic dark matter~\cite{Abgrall2016b,Abgrall2017a}, and lightly ionizing particles~\cite{Alvis2018}. An important low-energy background in Ge detectors is caused by spallation reactions on Ge by high-energy cosmic neutrons at the earth's surface. The important case of \nuc{68}{Ge} production yields in enriched Ge was measured in Ref.~\cite{Elliott2010}.  The isotope \nuc{68}{Ge} is removed only at the enrichment stage, but both zone refining and crystal growth remove all other cosmogenic isotopes. Hence, surface exposure after each of these steps is a concern. This exposure was addressed for the \MJ\ \DEM\ detectors in several ways. First, the enriched GeO$_2$ was shipped from Russia in a steel shipping container, developed by GERDA, that reduced the cosmogenic production of \nuc{68}{Ge} by a factor of approximately 10. In addition, a zone-refining facility was established adjacent to the ORTEC, Inc. detector production facility and a ten-minute drive from the Cherokee Caverns, which allowed convenient underground storage of the Ge between processing steps. Finally, each part was tracked through its history with a detailed database~\cite{Abgrall2015}. These procedures resulted in significant reductions in the low energy background, especially tritium $\beta$-decay, and opened the door to  searching for other physics.  Although GERDA did not pursue a low-energy program, the collaboration followed a similar strategy to reduce cosmogenic backgrounds impacting \BBz.

\subsection{GERDA}
The GERmanium Detector Array (GERDA) for \nuc{76}{Ge} experiment arose from the idea of using liquid nitrogen (LN) as a shield because of its low radioactivity. The idea, originated by Heusser~\cite{Heusser1995}, was to immerse bare Ge detectors in LN, which would act as coolant and shield. This concept was developed by the GErmanium in liquid NItrogen Underground Setup (GENIUS) collaboration~\cite{Klapdor2003a} and realized by GERDA. The GERDA collaboration~\cite{Ackermann2013,Agostini2013a,Agostini2017,Agostini2018}, however, used liquid argon (LAr) instead of LN due to its higher $\gamma$-ray stopping power. In addition, the LAr is an excellent scintillator, and was very effective as veto against background radiation external to the detector array itself.The initial GERDA goal was to confirm or refute the claim for the observation of \BBz~\cite{Klapdor2004b,Klapdor2006}.

The Ge detectors in GERDA are deployed in 7 strings, each enclosed within a nylon shroud that prevents radioactive ions (\nuc{42}{K} in particular) from electrostatic attraction to the detector surface. The group of strings is submerged in a 64 m$^3$ volume of LAr. The cryostat containing the LAr is, itself, contained within a 590~m$^3$ volume of pure-water. The neck of the LAr cryostat provides access for, not only the detectors, but all the associated utilities and data acquisition readout. The experiment is running at LNGS at a depth of 3400~mwe.

\begin{figure}[h!]
\begin{center}
\includegraphics[width=8cm]{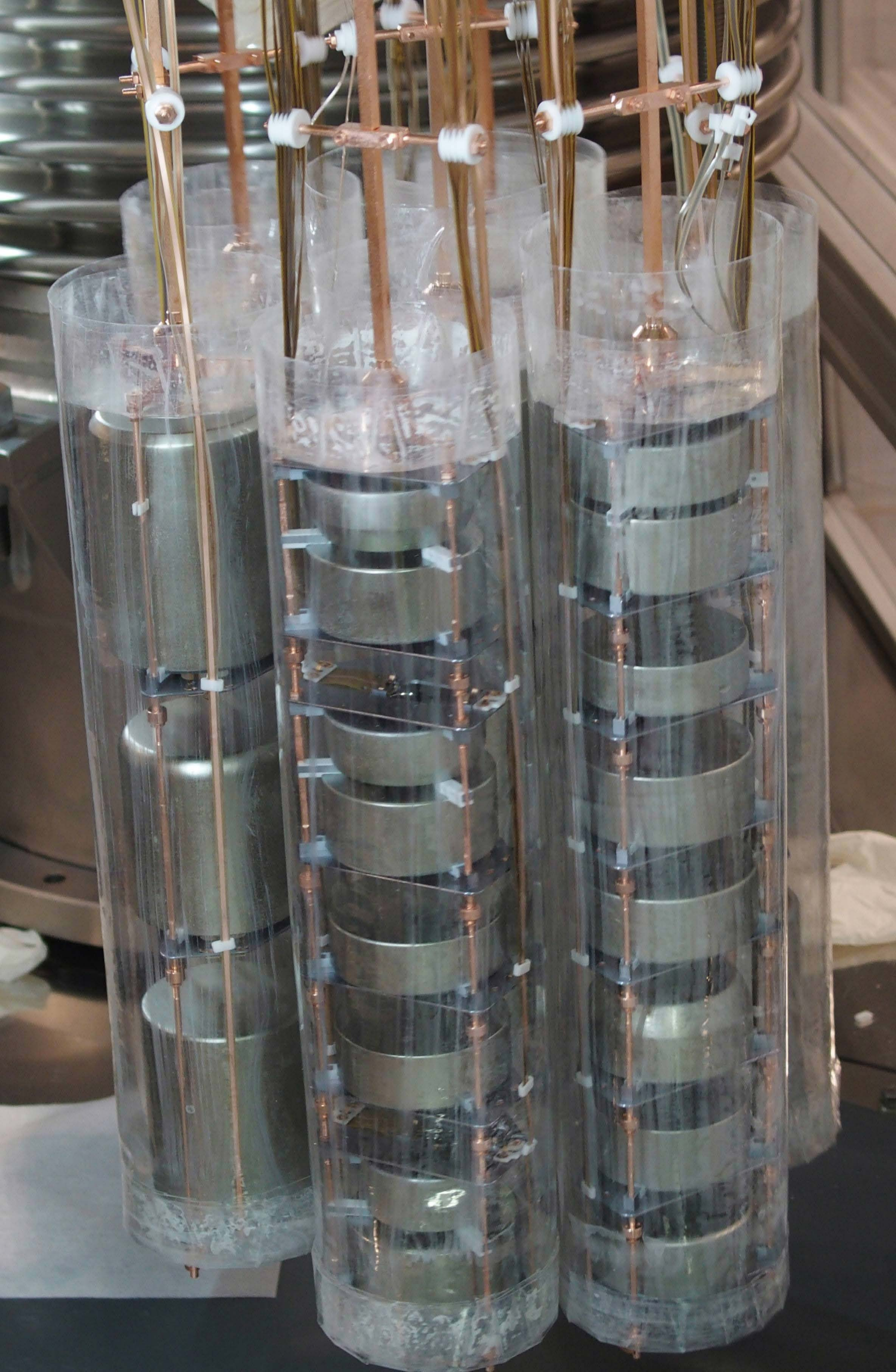}
\includegraphics[width=7cm]{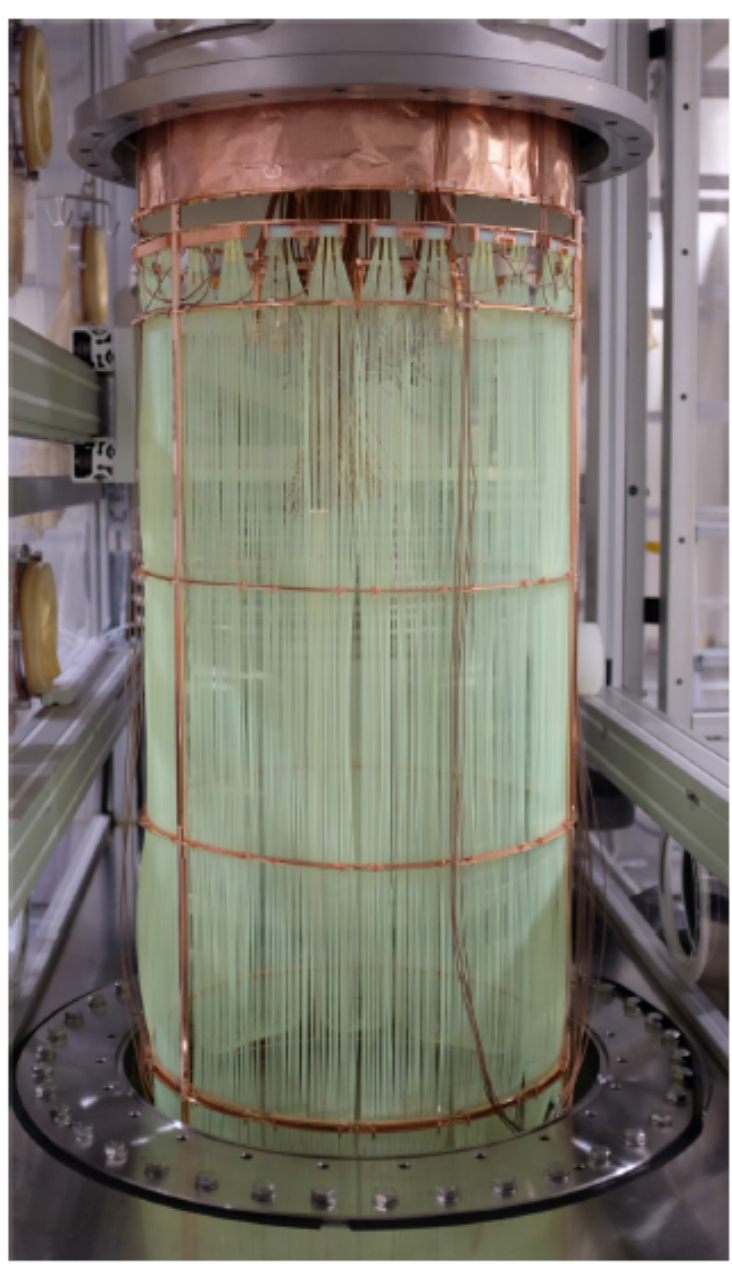}
\end{center}
\caption{ Left: A photograph of the strings of detectors for GERDA. Right: A photograph of the scintillation light collection system that surrounds the Ge detectors. Photos courtesy of the GERDA Collaboration.}\label{fig:GERDAstrings}
\end{figure}

The experiment has progressed through 2 phases. In Phase I, 17.6 kg of enriched Ge, including the detectors used by the HM and IGEX experiments, acquired 21.6 kg yr of data and found a half-life limit of \cpowten{2.1}{25} yr (90\% CL)~\cite{Agostini2013a}. The background index at \Qbb\ was 0.01 counts/(keV kg yr). Phase II increased the enriched detector mass to 35.6 kg and added a light detection system to the LAr surrounding the detectors. Figure~\ref{fig:GERDAstrings} shows the detector strings and LAr veto systems. This technique permitted a veto of events that deposited energy in both the Ge and Ar resulting in a significant background decrease to \cpowten{(5.6\pm3.4)}{-4} counts/(keV kg yr) in their BEGe dectectors~\cite{zsigmond2018}. This is the lowest background ever achieved by a \BBz\ experiment when normalized to the resolution at \Qbb. The reported combined exposure of Phases I and II is 82.4 kg yr resulting in a half-life limit of \cpowten{9.0}{25} yr (90\% CL)~\cite{Agostini2018}, convincingly ruling out the previous claim of \cpowten{(2.23^{+0.44}_{-0.31})}{25} yr~\cite{Klapdor2006}.  Ref.~\cite{Schwingenheuer2013} strongly criticizes this claimed value and argues that one should compare to the value in Ref.~\cite{Klapdor2004} of \cpowten{(0.69-4.18)}{25} yr with a quoted best value of \cpowten{1.19}{25} yr. At this time, both are excluded by the GERDA data. GERDA has also measured \Tt\ = \cpowten{(1.84^{+0.14}_{-0.10})}{21} yr~\cite{Agostini2013}, which was followed by \cpowten{(1.926\pm0.094)}{21} yr~\cite{Agostini2015}. Figures 2 and 3 in that latter paper shows a measured spectrum and fits including \BBt\ and the key background components. The dominance of \BBt\ is clear.

\section{LEGEND and the future of \BBz\ decay of \nuc{76}{Ge}}
When normalized to the resolution at \Qbb, GERDA has the lowest background of any \BBz\ experiment, with \MJ\ a close second. The two experiments have very modest exposures compared to other technologies but still have competitive or leading half-life limits. This situation has motivated the pursuit of a next-generation \BBz\ experiment based on \nuc{76}{Ge}. The Large Enriched Germanium Experiment for Neutrinoless Double Beta (\LEG) Collaboration~\cite{Abgrall2017d} aims to develop a phased, \nuc{76}{Ge} double-beta decay experimental program with discovery potential at a half-life beyond \powten{28} yr, starting with existing resources as appropriate to expedite physics results. This goal has led to a phased program, \Ltwo\ and \Lthou. \Ltwo\ will deploy up to 200 kg of Ge detectors within the existing GERDA infrastructure at LNGS. Only modest modifications to the lock at the top of the cryostat and the piping in the cryostat neck are required to accommodate the increased detector mass. In \MJ\ the components near the detectors, such as the front-ends and cables, were very radio-pure. In GERDA, the LAr veto was a very powerful tool for rejecting background. Using the more radio-pure parts and improving the light yield of the LAr veto system will reduce the background to 0.6 counts/(FWHM t yr) (\cpowten{2}{-4} counts/(keV kg yr)). The 3$\sigma$ discovery level for this configuration is estimated to be greater than \powten{27} yr. Figure~\ref{fig:DiscoveryPotential} shows the discovery potential of a Ge experiment as a function of exposure for several background levels. To reach the intended goal, \Ltwo\ requires about 1 t yr of exposure. The experiment is anticipated to begin operations in 2021.

\Ltwo\ is nearly fully funded with a few requests still pending. The project is under development at the time of this writing. \Lthou\ is envisioned to deploy a ton of isotope within 5 payloads into LAr. (See Fig.~\ref{fig:L1000cryostat}.) The goal is to reach a limit of $>10^{28}$ yr.

\section{Conclusion}
Germanium detectors have excellent energy resolution and very low background. As a result, limits on \Tz\ from Ge are very competitive even when the exposure is much less than competing technologies. Detectors fabricated from Ge have historically provided outstanding constraints on \Tz\ and \mee. From the first Ge-based experimental result in 1967, limits on \Tz\ have improved by a factor of \cpowten{2}{5} over the intervening 50 year period. The technology continues to advance and an additional improvement in sensitivity of more than a factor of 100 is within reach in the near future.

\section*{Conflict of Interest Statement}

The authors declare that the research was conducted in the absence of any commercial or financial relationships that could be construed as a potential conflict of interest.

\section*{Author Contributions}

FTA and SRE equally contributed to the preparation of this manuscript.  

\section*{Funding}
FTA thanks the National Science Foundation for support under grant NSF1614611. SRE thanks the Department of Energy Office of Nuclear Physics for support under contract number DE-AC52-06NA25396. SRE acknowledges the support of the U.S. Department of Energy through the LANL/LDRD Program.

\section*{Acknowledgments}
The authors would like to acknowledge the collaborations that provided input for this manuscript including \MJ, GERDA, and LEGEND. We specifically thank Bernhard Schwingenheuer, Riccardo Brugnera, and Vince Guiseppe for careful readings.

\bibliographystyle{frontiersinHLTH&FPHY} 
\bibliography{DoubleBetaDecay.bbl}


\section*{Figure captions}


\begin{figure}[h!]
\begin{center}
\includegraphics[width=12cm]{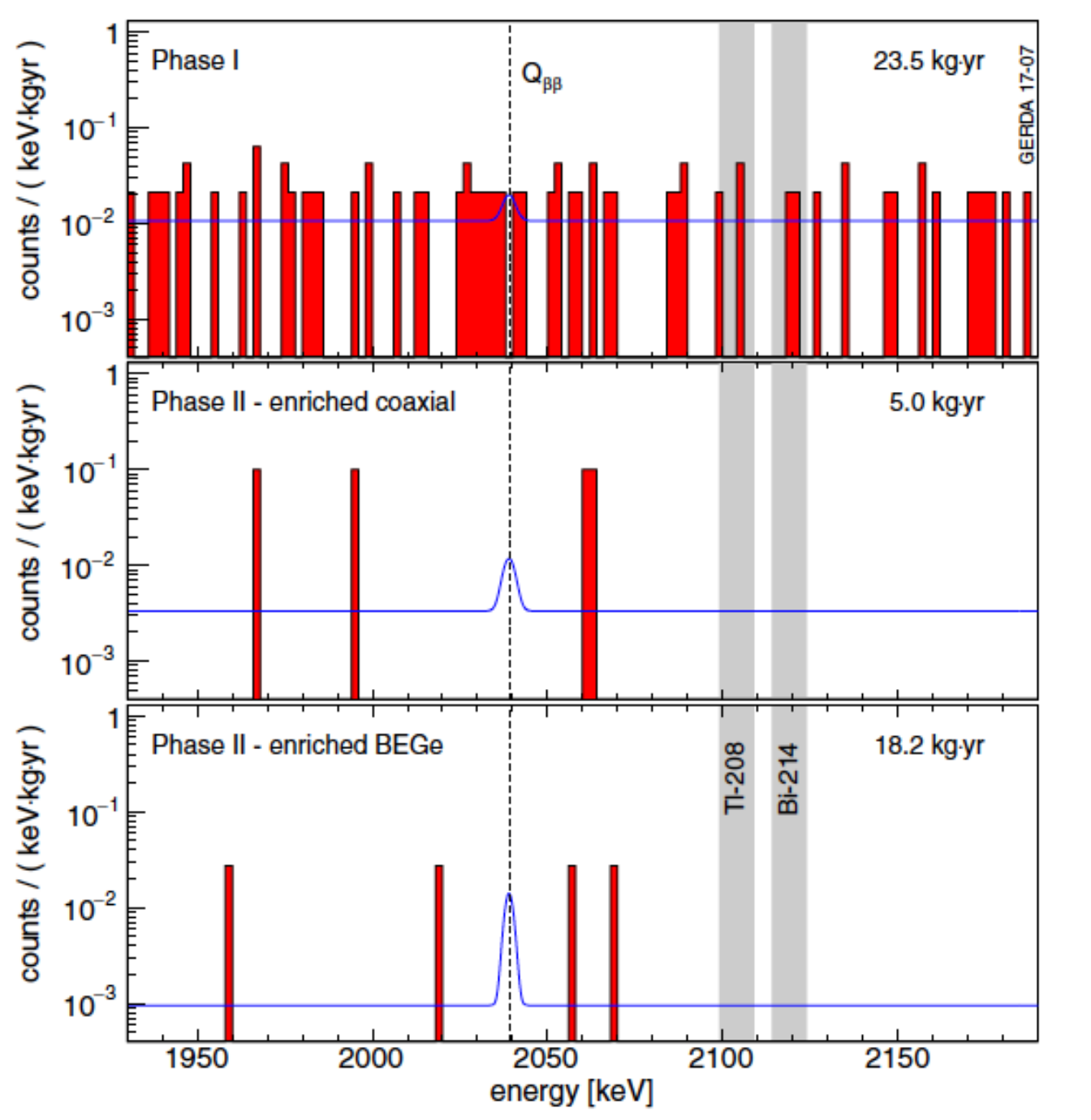}
\end{center}
\caption{The spectra from the various phases of GERDA near the \Qbb\ for Ge at 2039 keV~\cite{Agostini2018}. Figures courtesy of the GERDA Collaboration.}\label{fig:GERDAspectra}
\end{figure}

\begin{figure}[h!]
\begin{center}
\includegraphics[width=12cm]{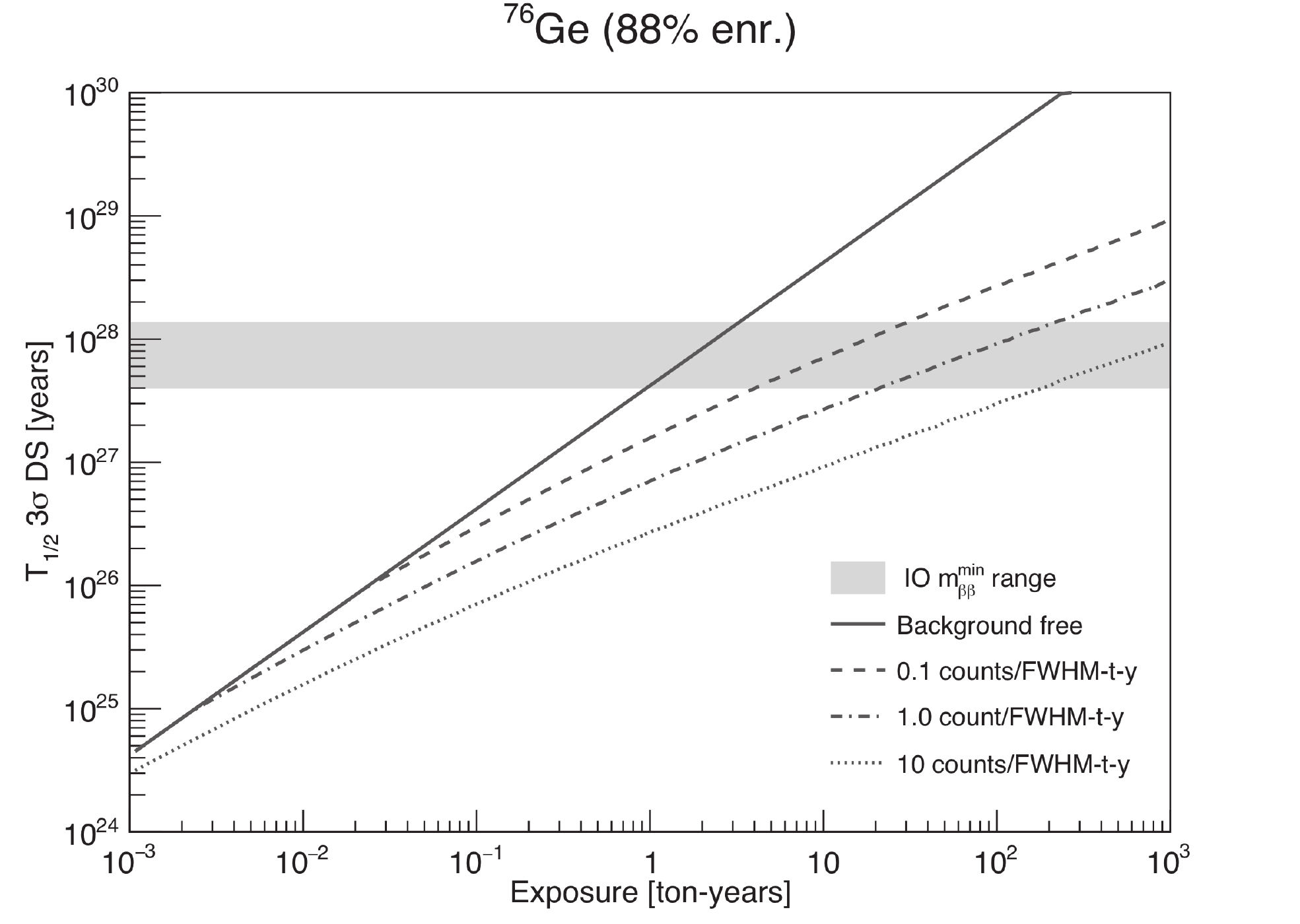}
\end{center}
\caption{The 3$\sigma$ discovery potential for a \nuc{76}{Ge} experiment for several potential background levels. Figure courtesy of Jason Detwiler.}\label{fig:DiscoveryPotential}
\end{figure}

\begin{figure}[h!]
\begin{center}
\includegraphics[width=12cm]{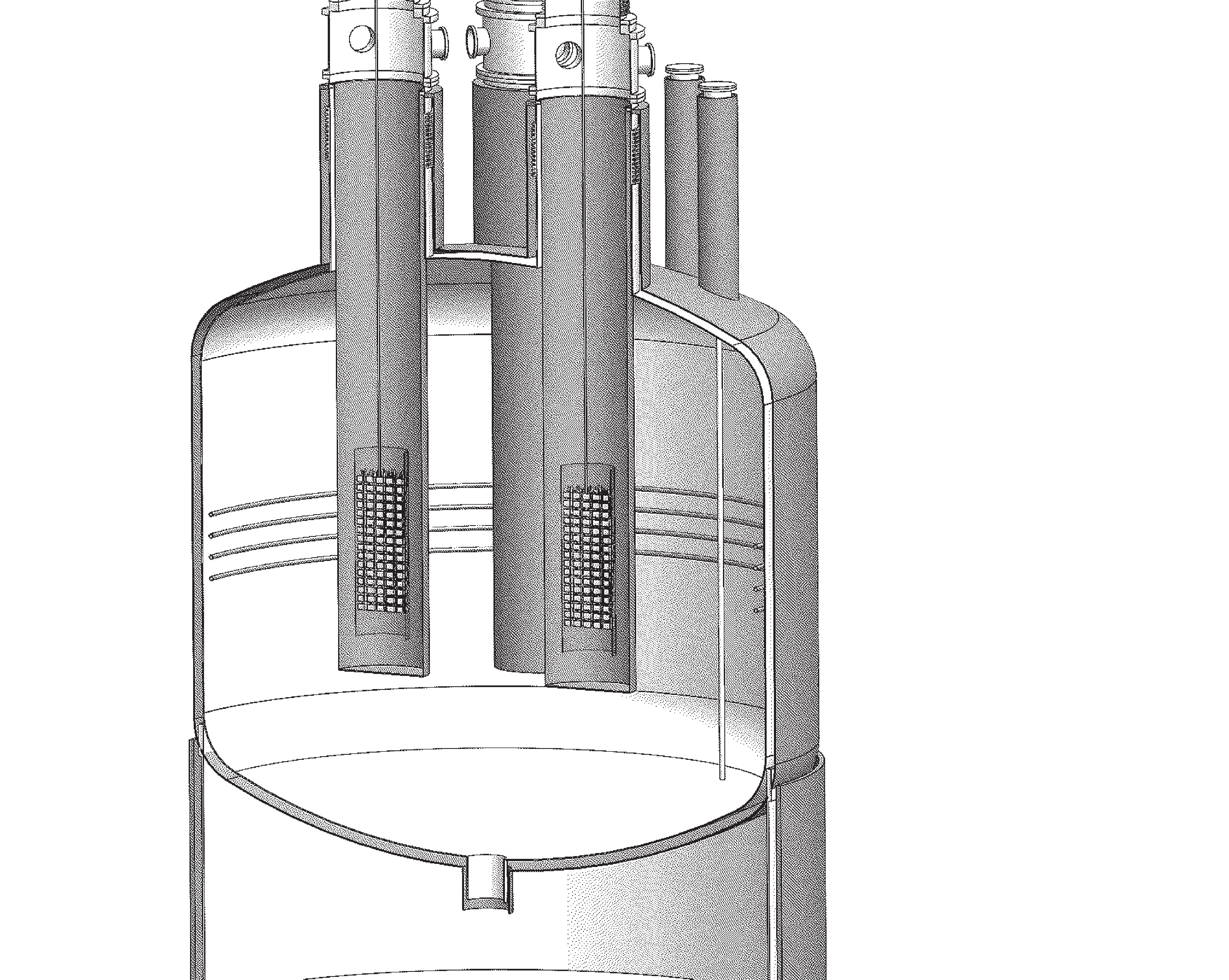}
\end{center}
\caption{The concept for \Lthou\ showing a number of the deployments Ge detectors.  This cut-away view shows three of five 200-kg groupings of Ge. Figure courtesy the LEGEND Collaboration.}\label{fig:L1000cryostat}
\end{figure}


\end{document}